\documentclass[12pt]{iopart}
\usepackage{graphicx}
\begin{document}


\title[Azimuthal and $\Delta\eta$ correlations with strange particles  
       at intermediate-p$_T$ at RHIC]{Azimuthal and pseudo-rapidity correlations with strange particles at intermediate-p$_T$ at RHIC}

\author{J Bielcikova for the STAR Collaboration}

\address{Physics Department, Yale University, New Haven, CT-06520, USA}
\ead{jana.bielcikova@yale.edu}
\begin{abstract}
 We present results on two-particle azimuthal correlations with strange trigger particles ($K^0_S$, $\Lambda$, $\Xi$, $\Omega$) associated with unidentified charged particles 
in d+Au and Au+Au collisions at $\sqrt{s_{NN}}$~=~200~GeV. We investigate, in detail, the near-side associated 
yield as a function of centrality, $p_T$ and strangeness content in the trigger particle 
to look for possible flavor and  baryon/meson differences. We compare our results to a fragmentation 
and recombination model, where the study of $\Omega$-triggered correlations is used as a critical test of the validity of the recombination picture. 
\end{abstract}



\section{Introduction}

The suppression of inclusive $p_T$ spectra of identified
particles in central Au+Au collisions with respect to p+p and
peripheral Au+Au collisions~\cite{Adler:2003kg,Adams:2003am} and
the enhanced baryon/meson ratios~\cite{Adams:2006wk,Abelev:2006jr} 
show that baryons and mesons behave differently than in p+p collisions for $p_T$~=~2-6~GeV/$c$. 
This indicates that fragmentation is not dominant and parton recombination 
and coalescence models~\cite{Fries:2003kq,Greco:2003mm,Greco:2003xt,Hwa:2002tu} 
have been suggested as alternative mechanisms of particle production.
Moreover, studies of di-hadron correlations in Au+Au
revealed the presence of an additional long-range pseudo-rapidity correlation ({\it ridge})
on the near-side~\cite{PutschkeQM}, which is absent in p+p and d+Au collisions. 

In this paper, we discuss the properties of two-particle correlations 
using strange trigger particles ($K^0_S$, $\Lambda$, $\Xi$, $\Omega$) 
associated with unidentified charged particles in d+Au and Au+Au collisions at 
 $\sqrt{s_{NN}}$~=~200~GeV with STAR.  We investigate the near-side associated 
yield as a function of centrality, system size, $p_T$ and strangeness content in the trigger particle, 
to look for possible flavor and  baryon/meson differences. We compare our results to a fragmentation 
and recombination model. 

\section{Data analysis}
The correlation functions, normalized to the number of trigger particles, are corrected 
for the reconstruction efficiency of associated particles, elliptic flow ($v_2$) 
and, unless mentioned otherwise, triangular acceptance in $\Delta\eta$. The near-side yield of associated 
particles is calculated as the area under the Gaussian peak obtained from a fit. We study separately 
the jet and ridge contributions to the near-side yield by analyzing the correlations in two 
$\Delta\eta$ windows: $|\Delta\eta|<$~0.7 containing both jet and ridge correlations, and $|\Delta\eta|>$~0.7
containing only the ridge contributions, assuming the jet contribution at large $\Delta\eta$ is negligible. 
The jet yield is free of systematic uncertainties due to the $v_2$ subtraction if a uniformity of $v_2$ with $\eta$ is assumed.
For the ridge yield, these systematic errors are estimated  by subtracting the $v_2$ measured by the event plane method (the lower bound) and by the 4-particle cumulant method (the upper bound). 

\section{Correlations with $K^0_S$, $\Lambda$ and charged trigger particles}
\begin{figure}[t!]
\begin{center}
\begin{tabular}{lr}
\includegraphics[height=6.5cm]{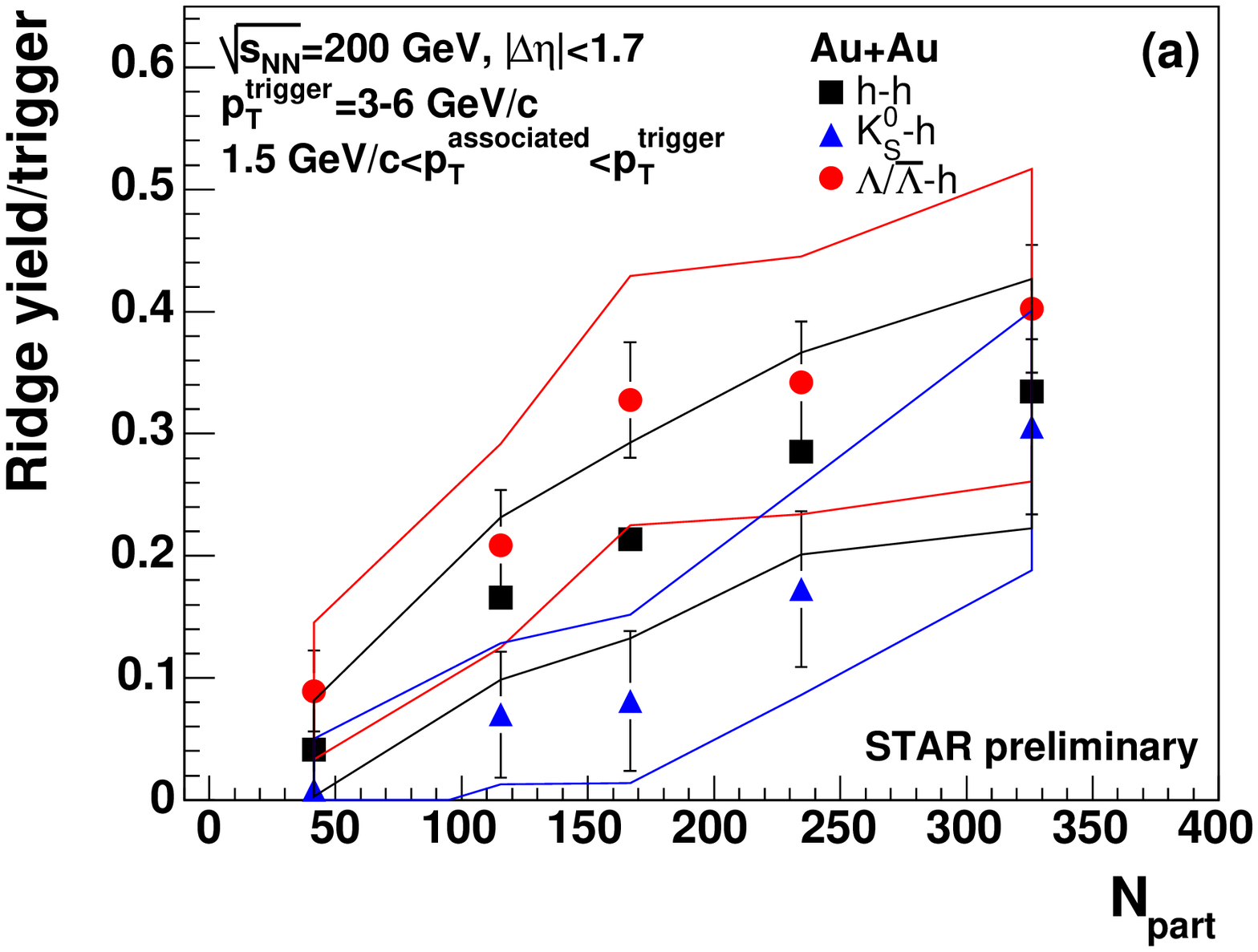}
&
\hspace{-0.5cm}
\includegraphics[height=6.5cm]{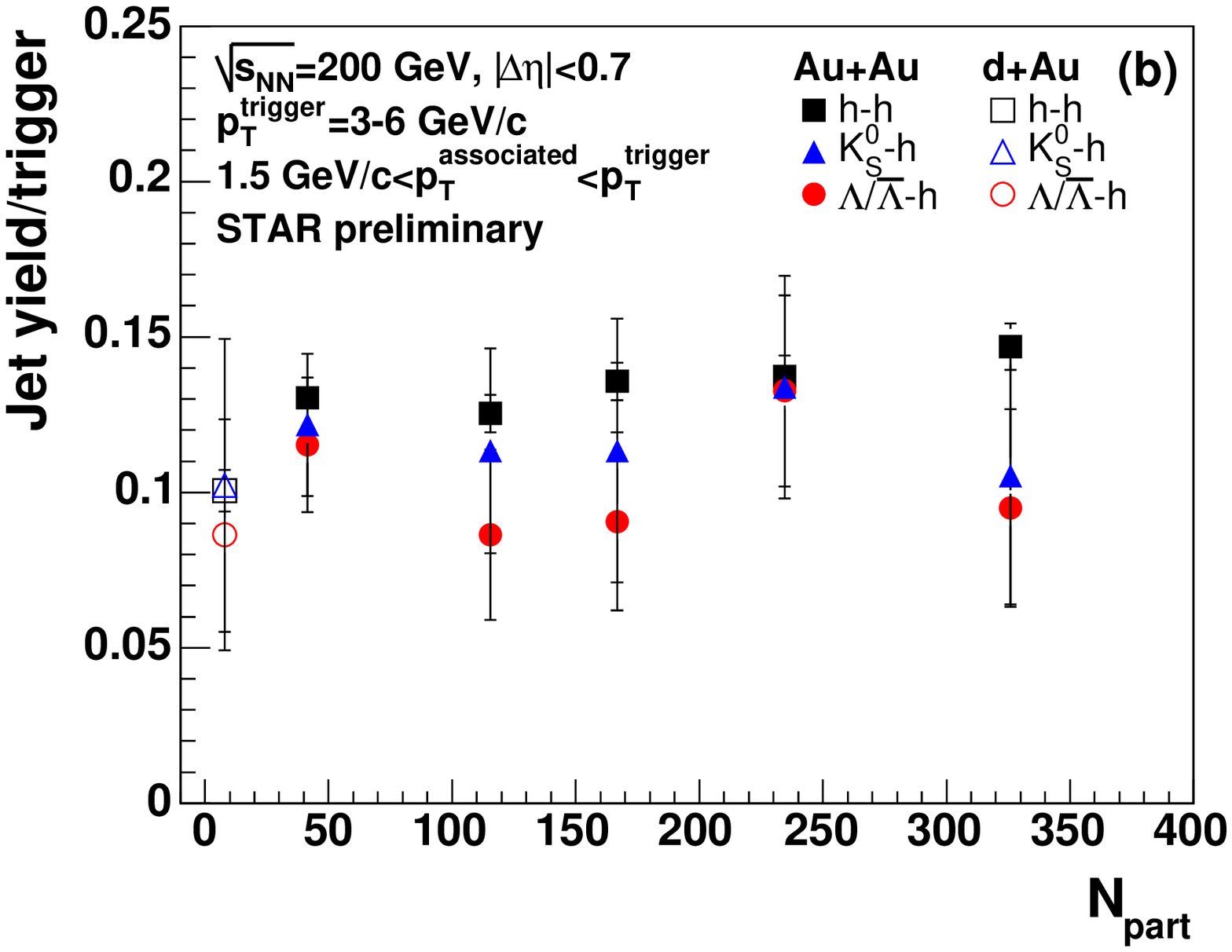}
\end{tabular}
\end{center}
\caption{Centrality dependence of the ridge yield (a) and jet yield (b) of associated charged particles for various trigger species in d+Au and Au+Au collisions. The error bands indicate systematic errors on the ridge yield due to the $v_2$ subtraction. } 
\label{rjcentr}
\end{figure}

Comparing the near-side yields in d+Au and Au+Au collisions, we observe 
a strong increase by a factor of 3-4 going from d+Au to central Au+Au collisions.
Studying separately the jet and ridge contributions to the near-side yield (Fig.~\ref{rjcentr}), 
we find that the ridge yield rises with centrality and is 
responsible for the observed strong increase of the near-side yield with centrality. 
The jet yield is independent of centrality and consistent with that in d+Au collisions. 
No significant baryon/meson or particle/anti-particle differences are observed.

\begin{figure}[t!]
\begin{center}
\begin{tabular}{lr}
\includegraphics[height=6.5cm]{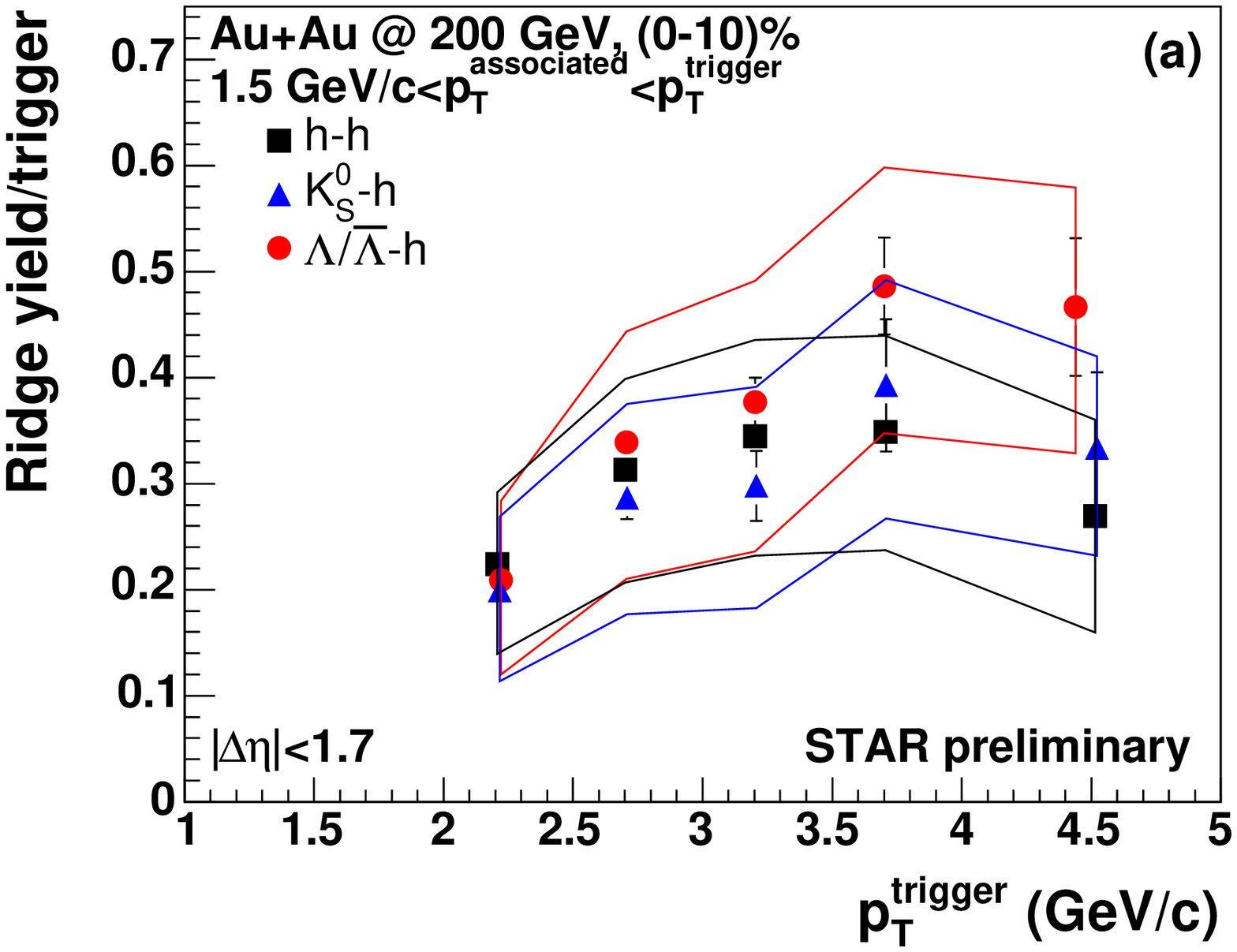}
&
\hspace{-0.5cm}
\includegraphics[height=6.5cm]{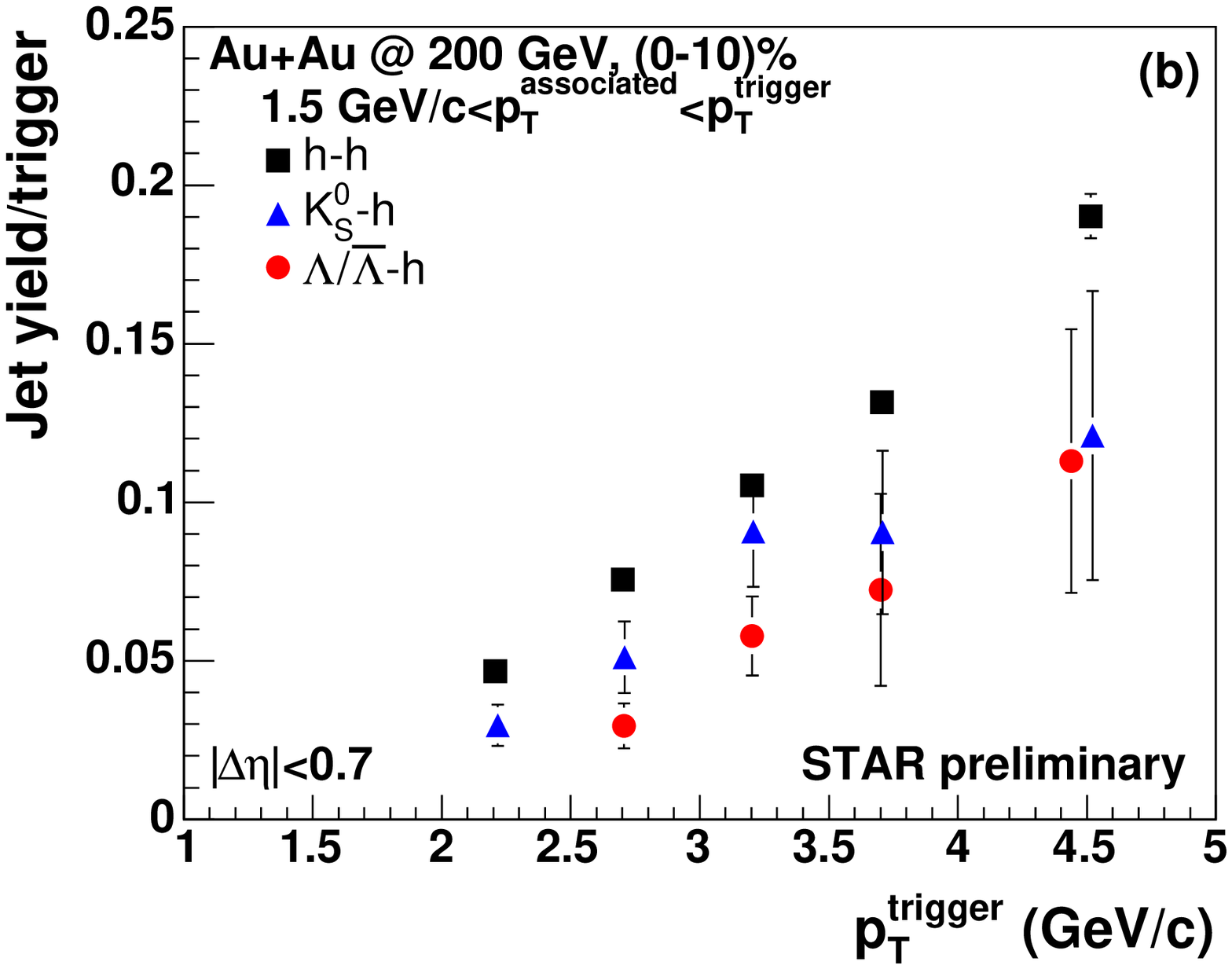}
\end{tabular}
\end{center}
\caption{Dependence of the ridge yield (a) and jet yield (b) on $p_T^{trig}$ for various trigger species in central (0-10\%) Au+Au collisions. The bands indicate systematic errors on the ridge yield due to the $v_2$ subtraction.}
\label{rjpttrig}
\end{figure}

Next, we study the dependence of the near-side yield on the transverse momentum of the trigger particle,
$p_T^{trig}$, shown in Fig.~\ref{rjpttrig}. While the ridge yield increases with $p_T^{trig}$ 
and flattens off for $p_T^{trig}>$~3.0~GeV/$c$, the jet yield
keeps increasing with $p_T^{trig}$ as expected. The jet yield for $\Lambda$ triggers
is systematically below that of charged hadron and $K^0_S$ triggers. Remaining 
effects of artificial track merging, which are found to affect more V0s than charged tracks, 
are currently under investigation.

We have also measured the invariant $p_T$ spectra of associated charged particles 
for $p_T^{trig}$~=~3-6~GeV/$c$ (not shown) and extracted the inverse 
slope, $T$. The ridge spectra have, for all studied trigger species, $T\sim$~400~MeV, 
close to that of 'the bulk', while 
the jet spectra have T$\sim$~450~MeV.  The  0-10\%/40-80\% centrality 
ratio of the near-side yields is  
about 3 at $p_T^{assoc}$~=~1~GeV/$c$ and decreases with $p_T^{assoc}$. As shown
in Fig.~\ref{rjcentr}, this large ratio is due to the correlations at large $|\Delta\eta|$ because 
the jet yield is independent of centrality. Our results qualitatively 
agree with the parton recombination model~\cite{Hwa:2005ui} which 
points toward a significant role of thermal-shower recombination 
in Au+Au collisions. To draw quantitative conclusions, the calculation must be done 
for the same centrality and $p_T^{trig}$ selection and also reproduce properties 
of the measured $\Delta\eta$ correlations.

\section{Correlations with multiply-strange trigger baryons}
Our study of correlations with multiply-strange baryons, especially $\Omega$, has been stimulated by the predictions from the recombination model of~\cite{Hwa:2006vb}. Contrary to the production of $K$ and $\Lambda$,  for particles created exclusively from strange quarks, such as $\phi$ and $\Omega$, the contribution from shower $s$ quarks should be negligible for $p_T$ up to 8~GeV/$c$. Consequently, there should be no associated particles for $\phi$- and $\Omega$-triggered correlations with $p_T>$~3~GeV/$c$ because they cannot be distinguished from background.

\begin{figure}[t!]
\begin{tabular}{lr}
\includegraphics[height=5.8cm]{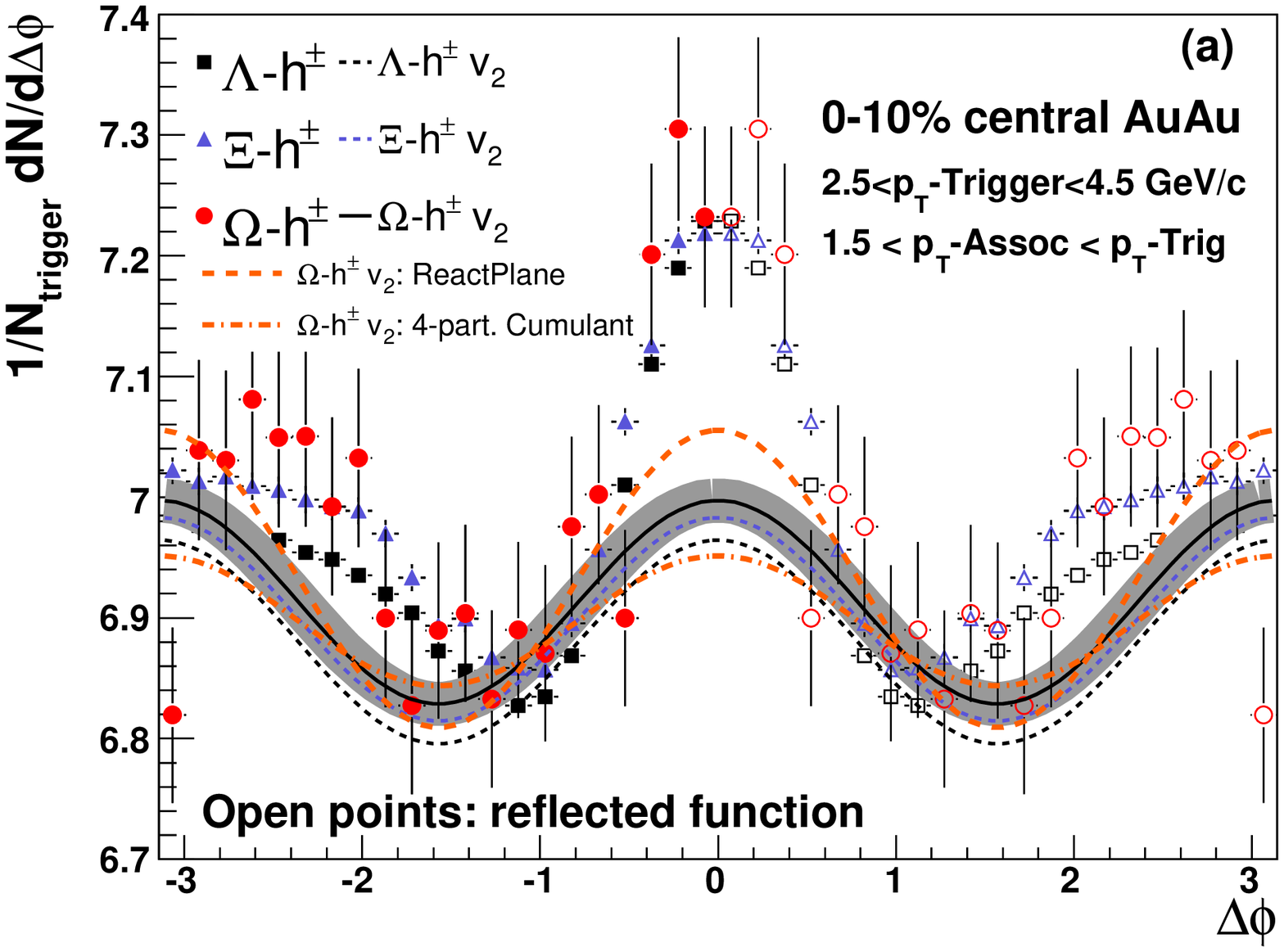}
\hspace{-0.5cm}
&
\includegraphics[height=6.3cm]{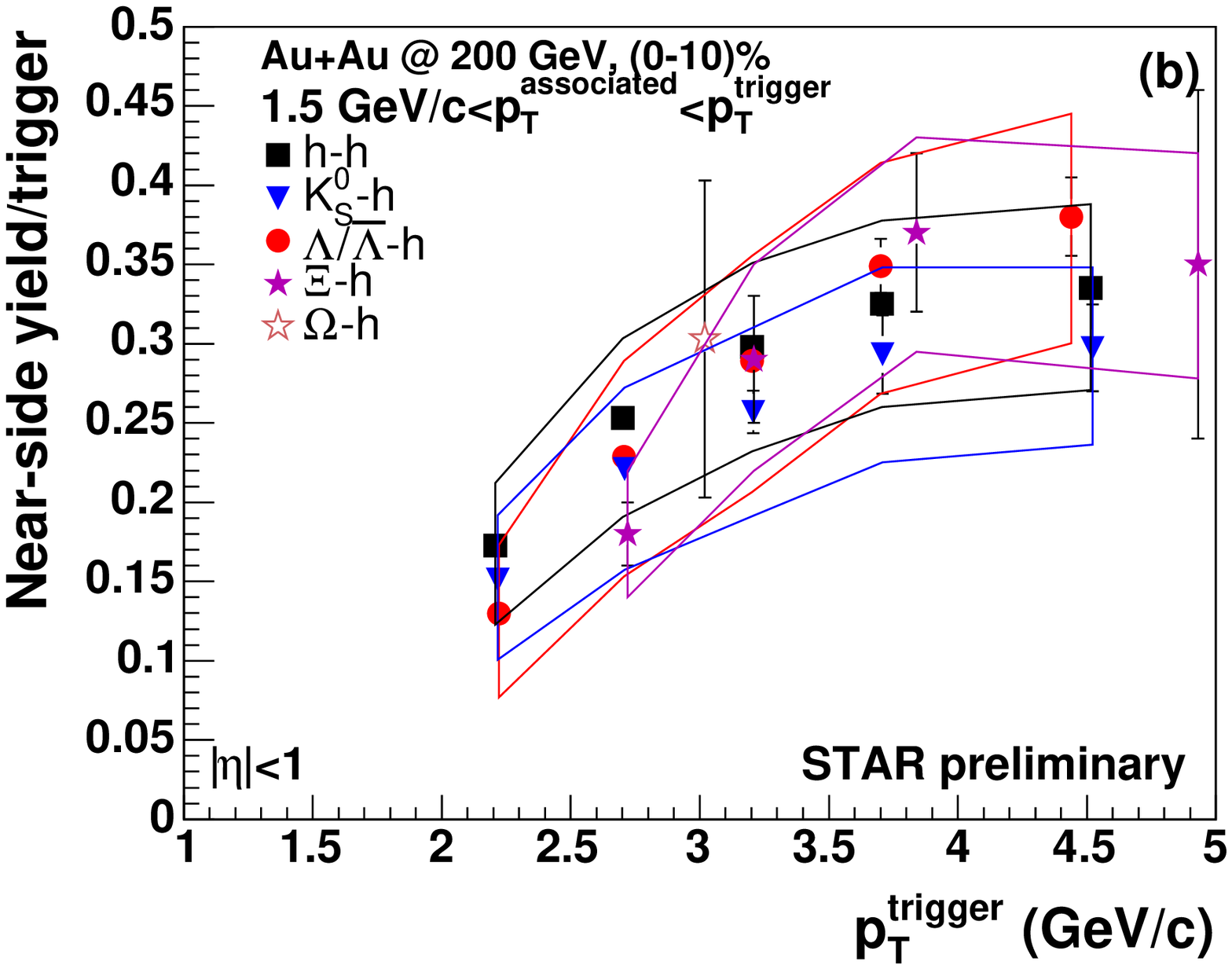}
\end{tabular}
\caption{(a) Azimuthal correlations with $\Lambda$, $\Xi$ and $\Omega$ for $p_T^{trig}$=2.5-4.5~GeV/$c$ in 0-10\% central Au+Au collisions.. 
The lines indicate the $v_2$ contributions. (b) Dependence of the near-side yield on $p_T^{trig}$ for various strange trigger particles in central (0-10\%) Au+Au collisions. The bands represent systematic errors due to the $v_2$ subtraction. No $\Delta\eta$ acceptance correction has been applied. } 
\label{lamomxi}
\end{figure}

Figure~\ref{lamomxi}(a) shows the azimuthal correlations for strange trigger baryons with 
increasing strangeness content: $\Lambda$, $\Xi$ and $\Omega$. Due to limited statistics, 
the trigger particles have been selected with $p_T^{trig}$~=~2.5-4.5~GeV/$c$. Clearly, there 
is a remaining near-side peak above the elliptic flow contribution for all discussed trigger species. The near-side peak persists even if the larger $v_2$, determined from the event plane method, is used.
Moreover, the strength of the near-side peak is, within errors, independent of the strangeness content in the trigger particle. This is further confirmed by the study of the $p_T^{trig}$ dependence of the near-side yield shown in Fig.~\ref{lamomxi}(b). Within errors, the yield is consistent for all studied baryon and meson species. 


\section{Conclusions}
We have reported results on two-particle correlations with strange trigger particles 
at intermediate-$p_T$ at RHIC. The correlations reveal a strong contribution 
from the long-range $\Delta\eta$ correlations in the near-side. We do not observe any significant
baryon/meson differences. The behavior of the central-to-peripheral ratio of the near-side yields 
agrees qualitatively with a recombination model. However, 
for  the $\Omega$-triggered correlations there exist associated particles in the near-side, 
which disagrees with the model. Studies of jet and ridge yields for $\Omega$-triggered correlations as well as correlations with identified associated particles are expected to constrain the origin of these correlations.

\end{document}